# Atomic-scale origin of charge density wave–driven metal–semiconductor transition in an incommensurately modulated metal-organic framework

Ling Zhang[1,‡], Zeyue Zhang[2,‡], Liu He[1], Bin Jiang[1,3], Yingchao Wang[4,5], Jiaxiang Zhang[1], Huimin Qi[1], Chao Zhang[6], Jinkun Guo[1], Hao Chen[1], Yunlong Fan[1], Yanran Shen[2], Hongli Jia[7], Guobao Li[2], Yu-Qing Zheng[3], Julius J. Oppenheim[8], Tianyang Chen[9], Jian Wang[10], Lei Sun[3,4,11*], Junliang Sun[2*] and Jin-Hu Dou[1*]

[1]National Key Laboratory of Advanced Micro and Nano Manufacture Technology; Key Laboratory of Polymer Chemistry and Physics of Ministry of Education, School of Materials Science and Engineering, Peking University, Beijing, 100871, China.
[2]College of Chemistry and Molecular Engineering, Beijing National Laboratory for Molecular Sciences, Peking University, Beijing, China.
[3]Beijing Advanced Innovation Center for Integrated Circuits, School of Integrated Circuits, Peking University, Beijing 100871, China.
[4]Department of Chemistry, School of Science and Research Center for Industries of the Future, Westlake University, 600 Dunyu Road, Hangzhou, Zhejiang Province, 310030, China.
[5]Institute of Natural Sciences, Westlake Institute for Advanced Study, 18 Shilongshan Road, Hangzhou, Zhejiang Province, 310024, China.
[6]Instrumentation and Service Center for Physical Sciences, Westlake University, Hangzhou 310030, Zhejiang Province, China.
[7]State Key Laboratory of Natural and Biomimetic Drugs School of Pharmaceutical Sciences, Peking University, No. 38 Xueyuan Road, Beijing 100191, China.
[8]Department of Chemistry, Princeton University, Washington Road, Princeton, New Jersey, 08540, United States.
[9]School of Science and Engineering, The Chinese University of Hong Kong (Shenzhen), Longgang, Shenzhen, Guangdong, 518172, China.
[10]International Center for Quantum Materials, School of Physics, Peking University, Beijing 100871, China.
[11]Department of Physics, School of Science and Research Center for Industries of the Future, Westlake University, 600 Dunyu Road, Hangzhou, Zhejiang Province, 310030, China.
‡These authors contribute equally.
*Correspondence: doujinhu@pku.edu.cn (J.-H. D.)
                 junliang.sun@pku.edu.cn (J. S.)
                 sunlei@westlake.edu.cn (L. S.)

**Abstract**

The intrinsic incommensurate charge density wave in metal-organic frameworks has remained elusive due to the lack of direct evidence linking atomic-scale structural modulation to macroscopic electronic properties. Using high-quality $Pr_3HHTP_2$ (HHTP = 2,3,6,7,10,11-hexahydroxytriphenylene) single crystals as a model system, we precisely resolve, for the first time, the incommensurately modulated structure of a conductive metal-organic framework at 100 K (modulation vector $q$ = 0.39143(12) $c^*$) via temperature-dependent single-crystal X-ray diffraction. The subsequent observation of a reversible metal–semiconductor transition around 350 K, which perfectly synchronizes with the disappearance of the structural modulation, provides convincing evidence for the electronic origin of the lattice distortion. Guest water molecules stabilize the modulated phase by synergistically regulating the relative rotation of the linkers and the interlayer spacing, thereby optimizing the inter-linker interactions. This work establishes a concrete experimental criterion for one-dimensional charge density wave in metal-organic frameworks and provides an ideal platform for probing coupled electronic–lattice modulations.

**Introduction**

Traditionally, conventional crystals are featured as long-range periodic structures, forming uniformly spaced lattices in three-dimensional direct and reciprocal space (Scheme 1a, d)[1]. On this basis, superimposing an additional periodic perturbation onto the underlying lattice could generate a modulated structure and cause satellite reflections in reciprocal space. The modulation is categorized based on the ratio between the perturbation period and the fundamental lattice period: a rational ratio results in a commensurate modulation (Scheme 1b, e), whereas an irrational ratio results in an incommensurate modulation, also referred to as an aperiodic structure (Scheme 1c, f). In this unique ordered state, although the atomic arrangement lacks simple spatial periodicity, it maintains high determinism and long-range correlation[2-4]. In low-dimensional condensed matter physics, such a phenomenon is not merely a geometric break in symmetry but also serves as a direct structural precursor to complex correlated electronic states, such as charge density wave (CDW)[5-11]. The period of modulation determines transport properties since the one-dimensional (1D) incommensurate modulation should lead to Peierls-distorted band structures[12,13]. Detailed descriptions of incommensurate modulation would reveal insights into CDW, deepening the understanding of electrical properties of functional materials.

Metal-organic frameworks (MOFs) have recently emerged as promising candidates for exploring quantum states of matter[14]. Benefiting from their highly tunable structural characteristics, MOFs provide an ideal physical platform for in-depth investigations of incommensurate modulation and complex correlated electronic states[15-20]. Although tens of thousands of MOFs structures have been reported, accounts of intrinsic aperiodic structures remain exceedingly rare[21]. Most reported superlattices or incommensurate structures in MOFs originate from host-guest composite systems[20,22,23]. In these materials, the incommensurate modulation stems from the mismatch between the spatial arrangement of guest molecules and the host framework[21]. The driving forces for such phenomena are primarily space-filling and weak host-guest interactions rather than the ordering of the framework's intrinsic electronic states. To date, incommensurate modulation has been observed in only a handful of porous conductive framework materials. These phenomena are associated with exotic quantum states such as CDW[24-26] and spin-phonon polarons[27], and give rise to important physical properties including metallic conductivity[26], ultra-low thermal conductivity[28], and anomalous spin dynamics[27]. However, despite the expanding scope of research, a definitive correlation between the precise atomic structural evolution and the macroscopic electronic phase transition remains elusive, and an intrinsic metal–insulator/semiconductor transition directly corresponding to the structural transformation has yet to be observed. Consequently, the electronic origin of CDW in this system still requires a decisive experimental elaboration.

In this study, using high-quality $Pr_3HHTP_2$ (HHTP = 2,3,6,7,10,11-hexahydroxytriphenylene) single crystals as a model system, we established direct experimental evidence linking atomic-scale structural modulation to macroscopic electronic evolution. Notably, we provide the first precise determination of an incommensurately modulated structure in a conductive MOF. This lattice distortion and spontaneous symmetry breaking inherently reflect the modulation of the crystalline framework by electronic ordering within the CDW state. This correlation is further substantiated by variable-temperature transport measurements, where a reversible intrinsic metal–semiconductor transition at approximately 350 K synchronizes perfectly with the disappearance of the structural modulation. Furthermore, our results highlight the synergistic role of guest water molecules in stabilizing the modulated structure by simultaneously regulating the relative rotation of the organic linkers and the interlayer spacing, thereby optimizing the π-π stacking interactions. Overall, this study not only provides a substantial experimental for the emergence of 1D CDW in MOFs but also offers an ideal model system for exploring the coupling mechanisms between correlated electronic states and lattice modulations.

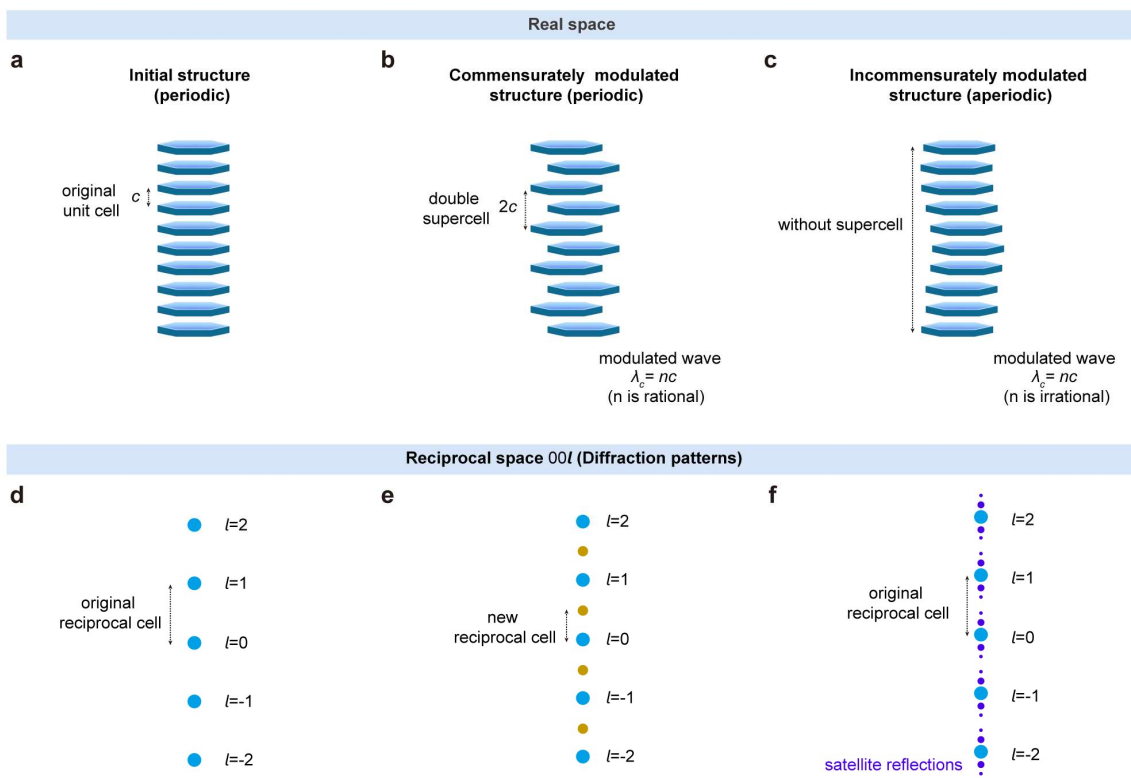

**Scheme 1 | Modulated structures and their diffraction patterns. a–c**, Real-space lattice models. **a**, Initial structure: a periodic lattice with the original unit cell ($c$). **b**, Commensurate modulation: a periodic lattice with a doubled supercell ($2c$), where the modulation wavelength is a rational multiple of $c$. **c**, Incommensurate modulation: an aperiodic lattice without a supercell, where the modulation wavelength is an irrational multiple of $c$. **d–f**, $00l$ diffractions in reciprocal space. **d**, Primary reflections at integer positions defining the original reciprocal cell (blue spots). **e**, Additional satellite reflections (brown spots) appearing at fractional positions (1/2) of the initial reciprocal unit cell. **f**, New satellite reflections (purple spots) emerging at non-integer positions flanking the primary peaks.

### Synthesis and characterization

Single crystals of $Pr_3HHTP_2$ were synthesized via a refined solvothermal protocol, in which the molar ratio of $Pr(NO_3)_3 \cdot 6H_2O$ to HHTP was optimized to 30:1 in a mixed solvent of $H_2O$ and $N,N$-dimethylacetamide (DMA). The synthesis yielded black, needle-like crystals exhibiting a pronounced metallic luster (Fig. 1h). To assess the synthetic universality and reproducibility, parallel experiments were conducted using HHTP ligands obtained from diverse sources, including commercial suppliers and those synthesized in-house. Nuclear magnetic resonance (NMR) spectra confirmed the high chemical purity of the ligands across all sources (Supplementary Figs. 1–3). All synthetic batches yielded single crystals with exceptional and consistent crystallinity regardless of the ligand source (Supplementary Figs. 4). Optical microscopy revealed highly uniform hexagonal rods (Fig. 1h) with lengths spanning 1–3 mm and diameters of 4–15 μm (Supplementary Figs. 5–7). Scanning electron microscopy (SEM) further corroborated the well-defined hexagonal prismatic morphology elongated along the crystallographic *c*-axis of the hexagonal lattice (Fig. 1h, inset). Additionally, high-resolution cryo-electron microscopy (cryo-EM) revealed clear lattice fringes with a measured spacing of approximately 19.2 Å (Fig. 1i). The corresponding Fast Fourier Transform (FFT) pattern generated by *CRISP* software[29] displayed sharp and intense spots, underscoring the long-range structural periodicity and high quality of the as-synthesized crystals (Fig. 1i, inset). These high-quality, large-scale single crystals provide an ideal platform for subsequent precise structural determination and intrinsic charge transport measurements.

The chemical composition and valence states were verified (detailed analytical results in the Supplementary text and Supplementary Figs. 8). Elemental analysis yielded an experimental stoichiometry of $Pr_{1.50}C_{18.56}H_{17.91}N_{0.36}O_{11.27}$, establishing a Pr-to-HHTP ratio of 1.5:1. The excess elemental content in the experimental formula is attributed to residual solvent molecules (e.g., DMA, water, and methanol) from the synthesis and washing procedures. Separately, X-ray photoelectron spectroscopy (XPS) analysis confirmed the exclusive presence of $Pr^{3+}$, with core-level peaks at 933.72 eV ($3d_{5/2}$) and 953.90 eV ($3d_{3/2}$), the absence of diagnostic features near 946/965 eV unequivocally rules out $Pr^{4+}$ in the framework[30-33]. Thermogravimetric analysis (TGA) revealed a stepwise weight loss of approximately 20% between 300 K and 473 K, which is primarily attributed to the removal of weakly physisorbed water/solvent molecules and more strongly bound guest species within the pores (Supplementary Figs. 9).

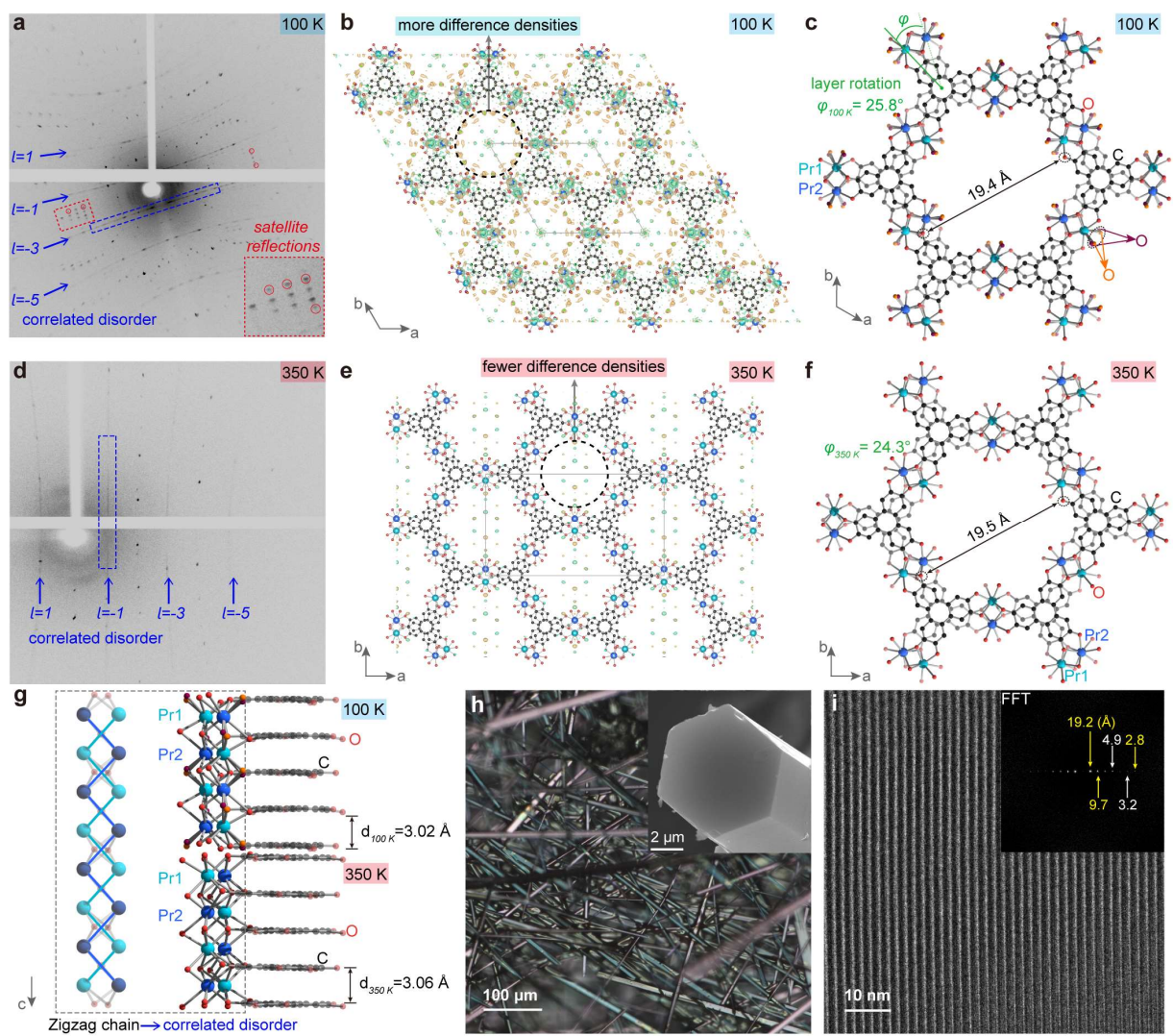

**Fig.1 | Structural and morphology characterization of Pr₃HHTP₂. a, d**, Single-crystal X-ray diffraction patterns at 100 K (**a**) and 350 K (**d**). **b, e**, Difference Fourier analysis indicating residual electron density within the channels at 100 K (**b**) and 350 K (**e**). **c, f**, Average crystal structures of Pr₃HHTP₂ viewed along the *c*-axis at 100 K (**c**) and 350 K (**f**). **g**, Interlayer spacing of the average structures and the zigzag chains of Pr³⁺ ions with mutually exclusive sites along the *c*-axis at 100 K and 350 K. **h**, Optical microscopy image of Pr₃HHTP₂ single crystals (inset: SEM micrograph of a hexagonal rod-shaped crystal elongated along the *c*- axis). **i**, High-resolution cryo-EM image of a Pr₃HHTP₂ rod normal to the *c*-axis (inset: corresponding Fast Fourier Transform (FFT) pattern).

**Structural analysis and incommensurate modulation**

Variable-temperature SCXRD data were collected for $Pr_3HHTP_2$ to analyze its crystal structure. This material retains high stability and crystallinity across the temperature range of 100 K to 400 K (Fig. 3a). Notably, a gradient phase transition was observed during the heating process: the average structure of $Pr_3HHTP_2$ was preserved at different temperatures while the satellite reflections corresponding to incommensurate modulation gradually disappeared during the heating process and vanished at ~350 K. Above 350 K, the material recovered a periodic structure.

The average crystal structures of $Pr_3HHTP_2$ under 100 K (incommensurately modulated) and 350 K (unmodulated) were solved and refined using the *JANA*2020 package[34] (Supplementary Table 1). Both structures share similar reticular structures with hexagonal pores whose diameters are 1.95 nm and 1.94 nm, respectively. These pore sizes are comparable to reported values of La- and Nd-based analogues[35]. The near-planar HHTP ligands lie within the *ab* plane and are stacked along the crystallographic *c*-axis. These stacks are interconnected by square-antiprismatic $Pr^{3+}$ ions between adjacent layers, resulting in a 3D **stp** topology (Supplementary Figs. 10). For each two neighboring HHTP ligands, the connecting $Pr^{3+}$ ion occupies two mutually exclusive sites with similar site energies, forming a zigzag chain along the crystallographic *c*-axis. Random arrangements of these zigzag chains at the two sites cause correlated disorder, which contributes to the diffuse scattering within the $l = 2n+1$ planes.

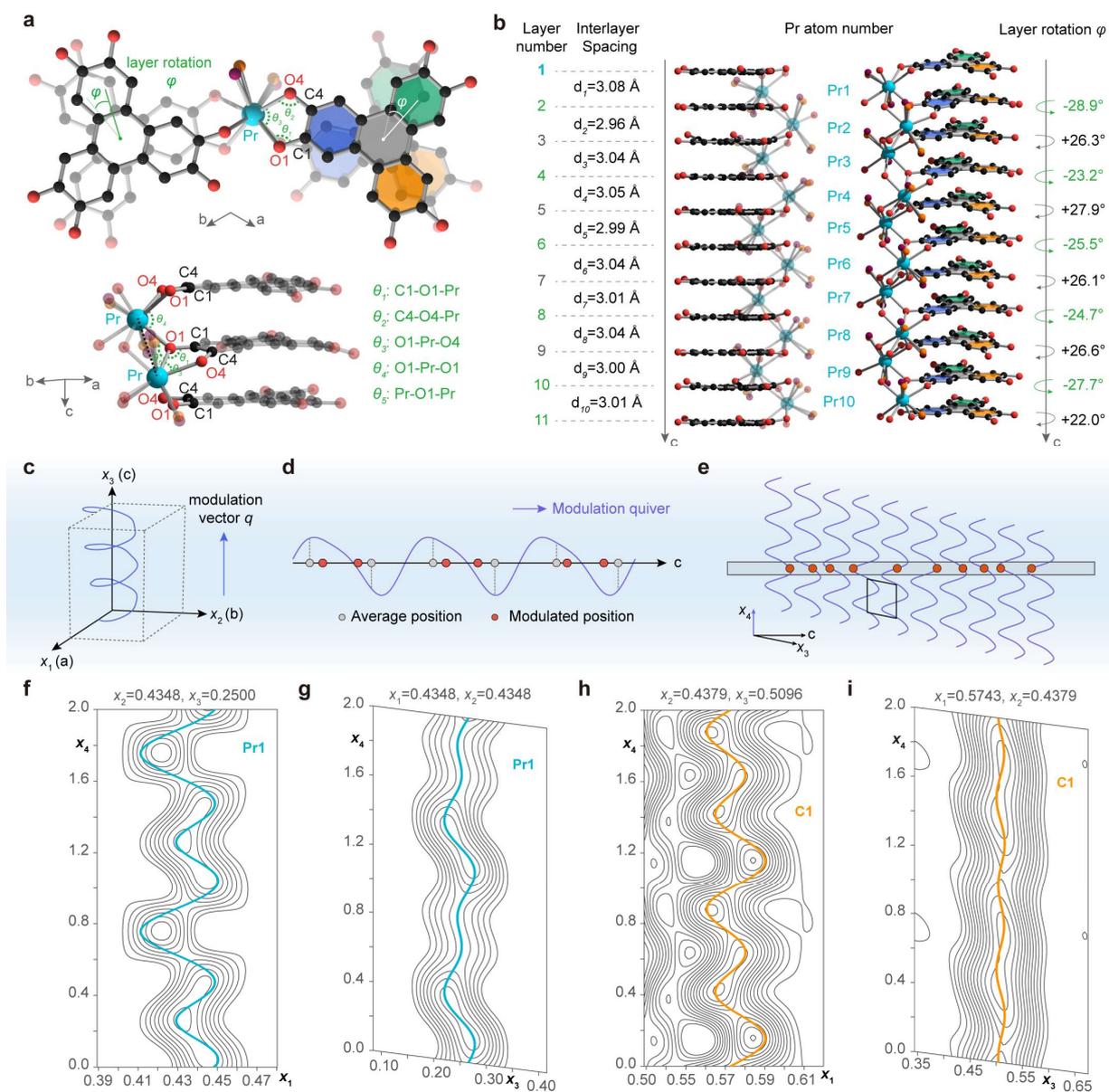

**Fig.2 | Incommensurately modulated structure and superspace analysis of Pr$_3$HHTP$_2$ at 100 K. a, b,** Approximate supercell models showing characteristic bond lengths, angles (**a**), and the evolution of interlayer distances and interplanar rotation angles (**b**). **c–e,** Schematic derivation of the (3+1) D superspace: **c.** visualization of the hypothetical modulation function in real space. **d,** Visualization of modulation function (blue), average positions (gray) and modulated positions (orange) along the $c$-axis. **e:** The $x_3$-$x_4$ section of the (3+1) D lattice, showing the unit cell, modulation functions, atomic positions and the definition of $c$, $x_3$ and $x_4$ axes. **g–i,** Electron density sections of the modulated dimension ($x_4$) plotted against spatial dimensions ($x_1$, $x_3$), representing the de Wolff sections for Pr (**f**: $x_1$-$x_4$, **g**: $x_3$-$x_4$) and C atoms (**h**: $x_1$-$x_4$, **i**: $x_3$-$x_4$).

We constructed and refined the incommensurate modulation model for $Pr_3HHTP_2$ using *JANA*2020 to elucidate the atomic-scale modulated structure at 100 K. While the structure of the $Pr_3HHTP_2$ after activation (where modulation and satellite reflections are lost) is already known, the structure of the as-synthesized phase—distinguished by its satellite reflections up to the second order—is reported here as a new and unusual structural state. According to the systematic absences concluded from the reciprocal sections shown in Supplementary Figs. 11, the space group of the average structure was determined to be $P\bar{3}c1$. Furthermore, the superspace group was determined to be $P\bar{3}c1(00\gamma)0s0$ with the modulation vector $q = 0.39143(12)\ c^*$. Notably, up to second-order satellite reflections can be observed, and these are significantly stronger than the first-order ones, which has not been observed in previous researches. Due to the appearance of additional satellites in the as-synthesized crystals that were previously been unseen, the stronger satellite reflections can be indexed as the second-order satellites of adjacent layers and the newly observed satellites are the first-order satellites. Correspondingly, the modulation vector is redefined as *q = (1-q')/2* where *q'* is the previously defined vector (here *q' = 0.21714 $c^*$*)[25,26]. A 1×1×5 (containing 10 layers) approximate supercell structure (see the Supplementary file) of $Pr_3HHTP_2$ at 100 K is displayed in Fig. 2a, b, showing that $Pr^{3+}$ and other atoms deviate from their average positions periodically. The approximate torsion angles of HHTP ligands on adjacent layers can be depicted as a function against the modulation coordinate *t*, as shown in Supplementary Figs. 12. For a more granular structural analysis, the specific bond lengths and bond angles within this incommensurate modulation framework are detailed in Supplementary Tables 2–4, providing a comprehensive quantitative description of the periodic lattice distortions. As shown in Fig. 2c–e, the periodic displacement of atoms can be depicted as a function of electron density in (3+1) D space, where sections of the modulated dimension $x_4$ against other dimensions are known as de Wolff sections. The de Wolff sections of several key atoms are displayed in Fig. 2f–i and Supplementary Figs. 13, which are the visualization of an equivalent representation of CDW.

The phase transition process manifests itself in the variable-temperature SCXRD data. Satellite reflections gradually disappeared during the heating process and all the satellite reflections vanished at approximately 350 K. (Fig. 3a). To evaluate the reversibility of this transition, a thermal cycle was conducted by heating the crystal to 360 K—ensuring the complete disappearance of the incommensurate modulation—followed by cooling back to 100 K (Fig. 3b). Although the modulated phase reappeared upon cooling, the final magnitudes of the *q* vector and the *a*/*c*-axis lattice parameters deviated from their initial states. The first-order satellites did not reappear upon cooling—only the second-order satellites reappeared—leading to a crystal that could be described either by $q = 0.39143\ c^*$ with mainly second-order satellites or by one described $q = 0.21714\ c^*$ for only first-order satellites. The irreversibility upon activation implies that the presence of solvents in the initial structure significantly perturbs the modulation. Difference Fourier analysis provides a structural rationale for changes in the structure upon cooling: clearly more residual densities are observed within the pores at 100 K, whereas these densities markedly decrease at 350 K (Fig. 1b, e), indicating the loss of guest $H_2O$ molecules during heating. Notably, the overall framework at 350 K remains consistent with the average structure at 100 K, despite minor symmetry-breaking disturbances. These findings suggest that the phase transition is primarily driven by temperature, while the extent to which the incommensurate modulation can be restored is closely correlated with the presence of guest $H_2O$ molecules.

To further scrutinize the decisive role of guest $H_2O$ molecules in preserving the incommensurate modulation, a controlled recovery experiment was performed on a single crystal of $Pr_3HHTP_2$ (Fig. 3c). Initial heating to 350 K induced the complete disappearance of satellite reflections. Subsequently, the crystal was re-exposed to ambient air followed by cooling to 100 K, this resulted in only a partial structural restoration, with the relative intensity of satellite reflections of 0.22847 and a modulation vector $q = 0.3776(4)\ c^*$, respectively. Remarkably, after immersing the same crystal in water for 8 h, SCXRD at 100 K revealed a significant enhancement in the relative intensities of satellite reflections to 0.31113, while the modulation vector $q$ increased to $0.3906(4)\ c^*$. This recovered state closely approaches the pristine characteristics observed at 100 K in Fig. 3a (intensity of satellite reflection: 0.37824; modulation vector $q$: 0.3842(5)) and Fig. 3b (intensity of satellite reflection: 0.39350; modulation vector $q$: 0.38790(19)), however without expression of the first-order satellites. Although absolute values exhibit minor fluctuations across different specimens due to intrinsic sample–to–sample variations, the consistent trend underscores that the extent of structural restoration is positively correlated with the degree of hydration—where direct immersion in water yields superior recovery compared to ambient air exposure. This strong correlation confirms that guest $H_2O$ molecules function as vital structural stabilizers that preserves the long-range order of the incommensurate modulation.

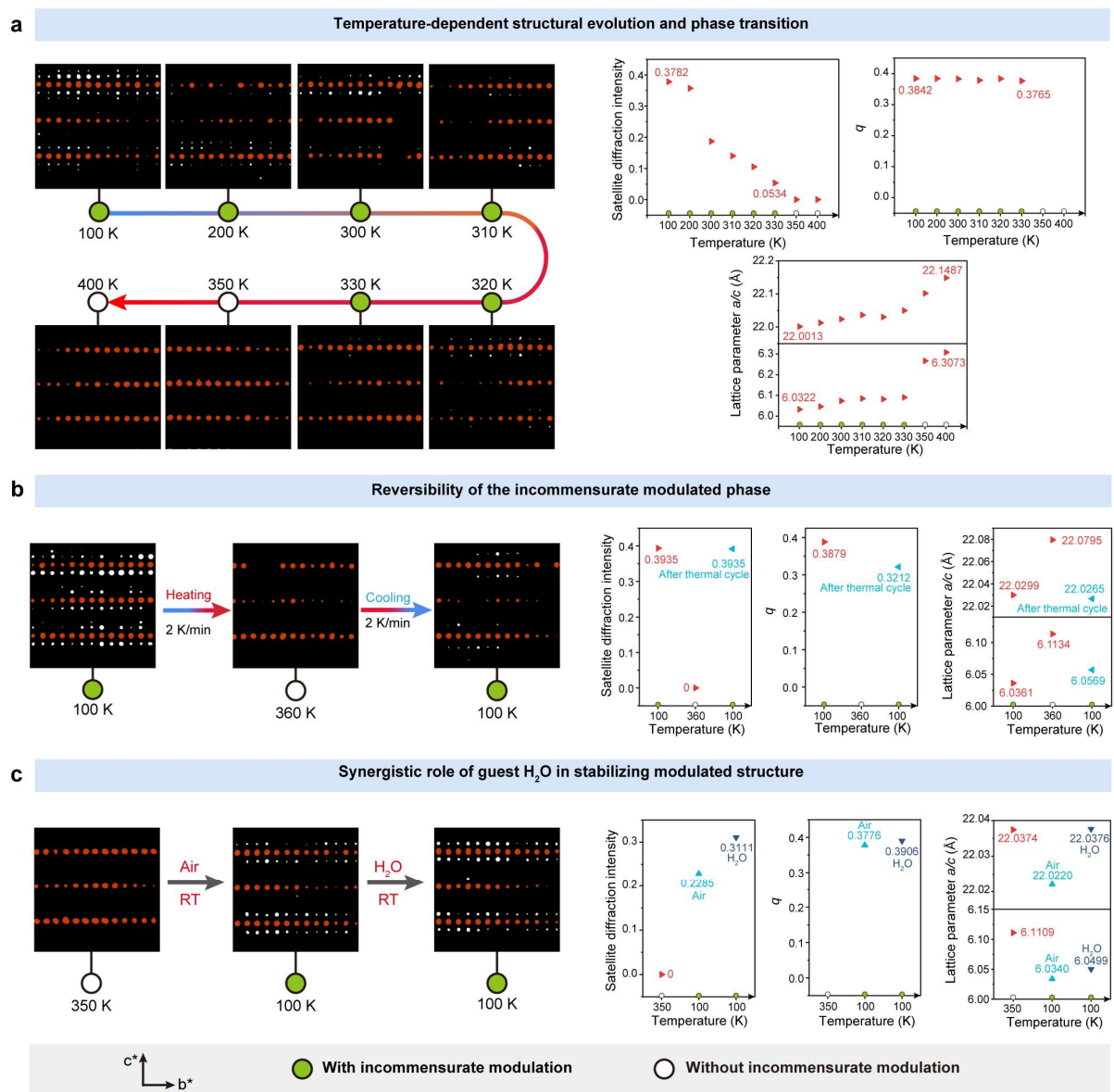

**Fig.3 | Phase transition and temperature-dependent modulation characteristics of Pr$_3$HHTP$_2$.** **a-c**, Comprehensive structural analysis under varied experimental conditions. **a**, Temperature-driven evolution of the crystal structure. **b**, Thermal reversibility tracking the disappearance and reappearance of the incommensurate modulation. **c**, Moisture-assisted recovery, investigating the synergistic role of H$_2$O molecules by monitoring structural restoration after air exposure and water immersion following high-temperature modulation loss. The subpanels in each group show (from left to right), respectively: (1) reciprocal space reconstructions projected along the [100] direction; (2) integrated intensities of the satellite reflections; (3) the modulation vector $q$; and (4) refined unit cell parameters ($a$, and $c$) as a function of temperature.

**Structure–dependent electronic phase transition**

The successful preparation of millimeter-scale hexagonal rod-like single crystals of $Pr_3HHTP_2$ (lengths up to 3 mm) provides a robust physical foundation for constructing highly reliable electronic devices. To ensure the structural integrity and retain the intrinsic transport properties of the material, four-probe devices were fabricated by precisely adhering single crystals onto pre-patterned gold electrodes using conductive paste (Supplementary Figs. 14). This direct contact method effectively circumvents potential lattice damage typically induced by invasive micro/nano-fabrication processes such as electron-beam lithography. Statistical analysis of 91 single-crystal devices from eight independent batches demonstrates exceptional synthetic reproducibility, with room-temperature electrical conductivities ranging from 117 to 883 S cm$^{-1}$ (Fig. 4a, Supplementary Table 5). Linear current–voltage (*I–V*) curves from different batches were observed in ambient conditions (Fig. 4b and Supplementary Figs. 15). We further evaluated the *I–V* characteristics under various environments—including ambient air, vacuum, and helium atmospheres—and across a wide range of temperatures (1.8–300 K). The *I–V* curves exhibited rigorous linearity under all conditions, confirming the Ohmic nature of the electrical contacts and ensuring that the observed transport phenomena reflect the intrinsic electronic properties of $Pr_3HHTP_2$ (Supplementary Figs. 16).

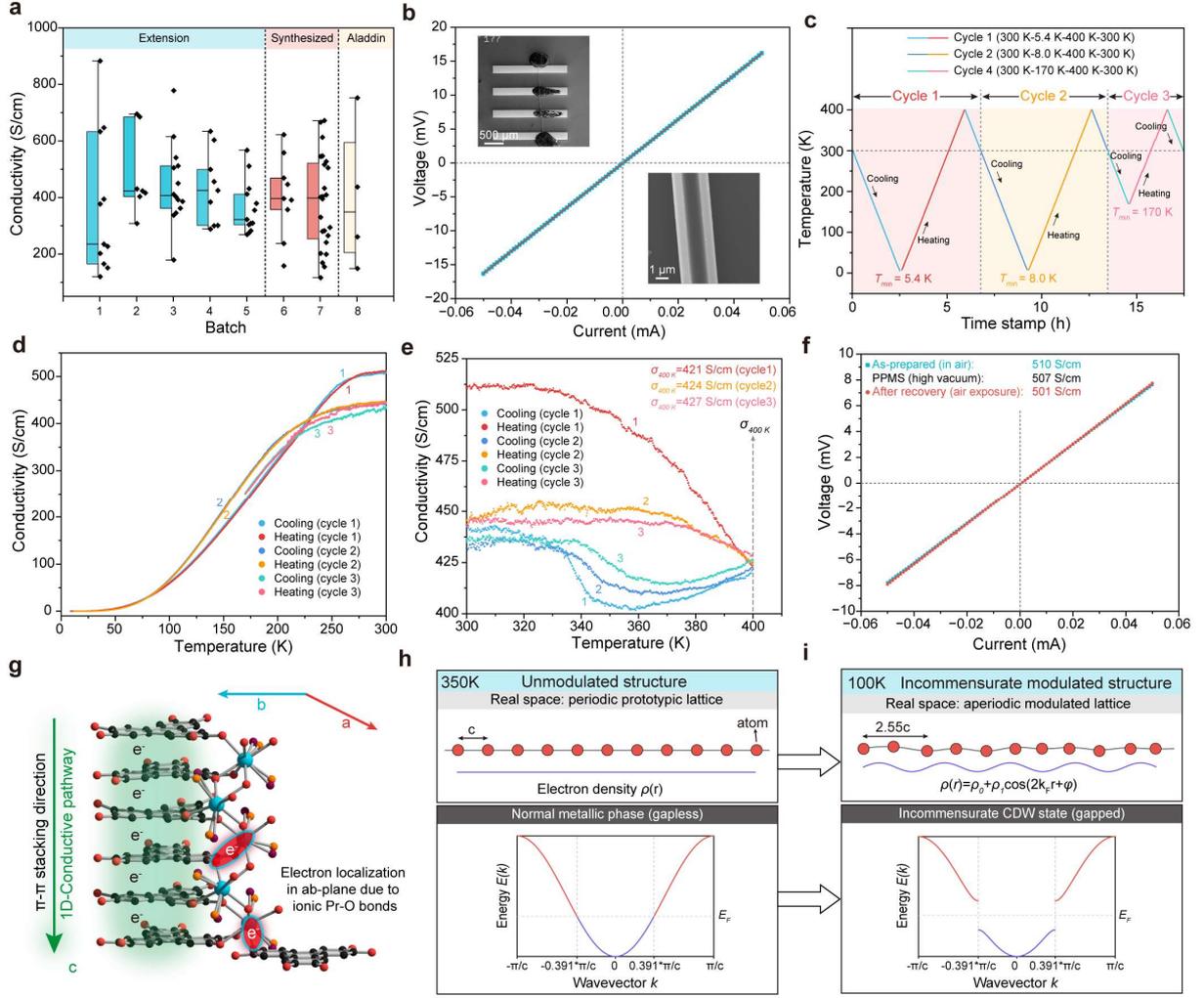

**Fig.4 | Electrical transport properties and 1D charge transport mechanism in Pr$_3$HHTP$_2$. a,** Room-temperature conductivity across eight independent batches. **b,** Representative linear *I–V* curve ($R^2 = 1$). Inset: SEM image of a single-crystal device. **c,** Temperature profile showing three consecutive heating–cooling cycles with varying minimum temperatures (scan rate: 2 K min$^{-1}$). **d, e,** Temperature-dependent conductivity over three consecutive cooling-heating cycles. **f,** *I–V* characteristic curves in ambient air before and 12 hours after PPMS cycles (following air exposure). **g,** Schematic of the 1D conductive channels formed by π-π stacking of HHTP ligands along the *c*-axis. **h, i,** Comparison of 1D electronic systems: (**h**) a periodic metallic lattice with uniform charge density, and (**i**) a Peierls-distorted insulating/semiconducting lattice with periodically modulated charge density and a Fermi-level energy gap.

To elucidate the intrinsic charge transport behaviors of $Pr_3HHTP_2$, we systematically investigated its electrical conductivity from 1.8 K to 400 K using a PPMS measurement setup (Supplementary Figs. 17). The transport properties were recorded over three heating-cooling cycles, and the corresponding heating-cooling curves of these three cycles are shown in Fig. 4c. Variable-temperature SCXRD characterization confirmed structural integrity within this temperature range (*vide supra*), and electrical characterization revealed a significant regulation of transport properties by thermal history. During the first heating from room temperature to 400 K, the conductivity dropped more sharply than in subsequent cycles. Differential scanning calorimetry analysis displayed a significant endothermic process upon heating the sample from room temperature to 373 K (Supplementary Figs. 18), confirming that the conductivity drop in the first cycle originated from the irreversible desorption of residual water molecules within the pores. Once the water loss reached saturation at 373 K upon continuous heating, the framework attained a stabilized state. The temperature dependence of electrical conductivity acquired in the second and third cycles almost overlapped, indicating that they reflect the intrinsic electrical properties of $Pr_3HHTP_2$ (Fig. 4d, e). Below 350 K, this material exhibits typical thermally activated semiconducting behavior. The stabilized activation energy ($E_a$) obtained from the Arrhenius fit ranges from 26.2 to 29.1 meV (based on fits of the second and third cycles), whereas in the first cycle, the $E_a$ values were lower and exhibited greater fluctuations (21.2–28.0 meV) due to the influence of residual moisture (Supplementary Figs. 19, 22, 23). Furthermore, these results are in agreement with the narrow optical bandgap of 0.78 eV derived from ultraviolet-visible diffuse reflectance spectroscopy and Tauc plot analysis at room temperature (Supplementary Figs. 20). The convergence of transport and optical data confirms the semiconducting nature of $Pr_3HHTP_2$. In contrast, above 350 K, the conductivity decreased with increasing temperature, displaying metallic characteristics. This electronic phase transition precisely synchronized with the disappearance of the structural modulation, indicating their shared origin. Finally, after re-exposure to ambient air for 12 hours, the cycled device recovered its original conductivity at 300 K: increasing from 436 S cm$^{-1}$ after the third cycle to 501 S cm$^{-1}$, which is nearly identical to the initial value of 510 S cm$^{-1}$ before thermal cycling (Fig. 4f). This reversibility not only demonstrates the robustness of the electrical conductivity but also reveals the synergy between temperature-driven structural modulation and water-assisted stabilization of the modulated structure. Furthermore, multiple independent devices show highly consistent transport characteristics (Supplementary Figs. 21–23). The metal–semiconductor transition temperature varies slightly depending on the specific device and thermal history, typically occurring within the range of 320 to 350 K. This experimental reproducibility confirms that the observed metal–semiconductor transition and thermal-cycle stability are intrinsic physical attributes of $Pr_3HHTP_2$.

Mechanistically, temperature governs the onset of the CDW phase and the opening of an electronic band gap via the Peierls transition. Meanwhile, guest $H_2O$ molecules within the pores play a crucial role in facilitating the recovery of the incommensurate modulated structure. By simultaneously regulating the relative rotation of the organic linkers and the interlayer spacing, these water molecules synergistically re-optimize the π-π stacking along the crystallographic *c*-axis through weak intermolecular interactions. This synergistic effect between the framework and guest molecules ensures high synchronicity and reversibility observed between electronic transport and lattice evolution. The physical essence of this system stems from its quasi-1D architecture, where HHTP ligands form highly oriented 1D conduction channels (Fig. 4g). In the unmodulated state above 350 K, the charge density is uniformly distributed without a bandgap, resulting in metallic behavior (Fig. 4h). However, according to Peierls theory[5,13,36], quasi-1D metals are inherently unstable due to strong electron-phonon coupling or electron correlation. Upon cooling, this instability induces spontaneous symmetry breaking, evolving into an incommensurate CDW state with space–modulated charge density. The experimentally observed modulation vector $q$ = 0.39143(12) $c^*$ confirms this incommensurability, where the collective mode opens an electronic band gap near the Fermi level, driving the intrinsic metal–semiconductor transition (Fig. 4i).

**Conclusion**

In summary, this study provides a comprehensive understanding of the 1D incommensurate modulation and its electronic origin in a MOF, $Pr_3HHTP_2$. By integrating temperature-dependent SCXRD with transport measurements, we have established decisive experimental evidence that links atomic-scale structural symmetry breaking to a macroscopic metal–semiconductor transition at approximately 350 K. The synchronization between the disappearance of the incommensurate modulation and the restoration of metallic behavior offers compelling evidence for a framework-driven CDW state. Furthermore, we have elucidated the synergistic role of guest water molecules in stabilizing this modulated state, where guest-mediated optimized π-π stacking is essential for sustaining long-range electronic correlations. These findings not only provide a solid criterion for identifying intrinsic aperiodic structures but also establish conductive MOFs as a versatile molecular platform for exploring the complex coupling between lattice dynamics and quantum states.

## Acknowledgements

We appreciate the Molecular Materials and Nanofabrication Laboratory (MMNL) in the College of Chemistry and Electron Microscopy Laboratory of Peking University for the support of instruments to perform device fabrication and characterization. We acknowledge State Key Laboratory of Natural and Biomimetic Drugs School of Pharmaceutical Sciences of Peking University for providing the instrumental support for SCXRD, and Fuling Yin for her valuable assistance during data collection. We thank the Instrumentation and Service Center for Material Sciences at Westlake University for the support with the electrical characterization. All the images were collected with FEI Titan Krios TEM (Gatan K3 summit camera) operated at 300 kV with GIF Quantum energy filter (Gatan). Cryo-EM data were collected on the Electron Microscopy Laboratory of Peking University with the assistance of Xuemei Li. We sincerely appreciate Tian Li, Guangfu Ni, and Shaoze Wang for helpful discussions.

## Author contributions

L. Z., L. S. and J. -H. D. designed the project and experiments. L. Z. and L. H. synthesized the samples. Z. Z. and J. S. performed the structural determination and analysis. L. Z. performed Scanning Electron Microscope (SEM), Optical microscope (OM) and UV-vis DRS measurement. B. J. performed TGA. H. Q. synthesized the HHTP ligand. J. G. performed cryo-EM characterization. L. Z., B. J. and H. J. collected the single-crystal data. L. Z and Y. W. fabricated single crystal devices and performed electrical characterization. L. Z. and C. Z. performed variable temperature-dependent electrical measurement. All authors participated in the writing of the manuscript.

**Funding**

This work was financially supported by the National Key R&D Program of China (Grant No. 2023YFE0206400), National Natural Science Foundation of China (Grant No. 22171185, 22575005), and the Clinical Medicine Plus X-Young Scholars Project at Peking University backed by the Fundamental Research Funds for the Central Universities. Y.W. and L.S acknowledge support from the Zhejiang Provincial Natural Science Foundation of China (XHD23B0301). The work also supported by the China Postdoctoral Science Foundation under Grant Number 2023M740055 and Postdoctoral Fellowship Program of CPSF under Grant Number GZC20230038.

**Competing interests**

The authors declare no competing interests.